\documentstyle[prl,floats,aps]{revtex}

\begin{document}

\input{epsf.sty}

\draft

\twocolumn[\hsize\textwidth\columnwidth\hsize\csname @twocolumnfalse\endcsname

\title{Gravitational Collapse of Gravitational Waves in 3D Numerical
Relativity}

\author{
Miguel Alcubierre${}^{(1)}$,
Gabrielle Allen${}^{(1)}$,
Bernd Br\"ugmann${}^{(1)}$,
Gerd Lanfermann${}^{(1)}$, \\
Edward Seidel${}^{(1,2)}$,
Wai-Mo Suen${}^{(3,4)}$, and
Malcolm Tobias${}^{(3)}$}

\address{
$^{(1)}$ Max-Planck-Institut f{\"u}r Gravitationsphysik,
Schlaatzweg 1, 14473 Potsdam, Germany
}
\address{
$^{(2)}$ National Center for Supercomputing Applications,
Beckman Institute, 405 N. Mathews Ave., Urbana, IL 61801
}
\address{
$^{(3)}$ Department of Physics,
Washington University, St. Louis, MO 63130
}
\address{$^{(4)}$ Physics Department,
Chinese University of Hong Kong,
Hong Kong}

\date{\today}
\maketitle

\begin{abstract}
  We demonstrate that evolutions of three-dimensional, strongly
  non-linear gravitational waves can be followed in numerical
  relativity, hence allowing many interesting studies of both
  fundamental and observational consequences.  We study the evolution
  of time-symmetric, axisymmetric {\it and} non-axisymmetric Brill
  waves, including waves so strong that they collapse to form black
  holes under their own self-gravity. The critical amplitude for black
  hole formation is determined. The gravitational waves emitted in the
  black hole formation process are compared to those emitted in the
  head-on collision of two Misner black holes.
\end{abstract}


\narrowtext

\vskip2pc]

Gravitational waves have been an important area of research in
Einstein's theory of gravity for years.  Einstein's equations are
nonlinear, and therefore can cause waves, which normally would
disperse if weak enough, to be held together by their own gravity.
This property characterizes Wheeler's geon~\cite{Wheeler55,Brill64}
proposed more than 40 years ago, and is responsible for many
interesting phenomena.  Even in planar symmetric spacetimes, there are
many interesting results, such as the formation of singularities from
colliding plane
waves (see \cite{Yurtsever88b} and references therein).  In
axisymmetry, Ref.~\cite{Abrahams92b} studied the formation of black
holes (BHs) by imploding gravitational waves, finding critical
behavior~\cite{Abrahams93a}.

These discoveries are all in spacetimes with special symmetries, but
they raise important questions about fully general
three-dimensional (3D) spacetimes, e.g.\ the nature of critical
phenomena in the absence of symmetries has only recently been studied
through a perturbative approach~\cite{Gundlach97b}. 3D studies of
fully nonlinear gravity can only be made through the machinery of
numerical relativity.  A few studies of gravitational wave evolutions
have been performed in the linear and near linear
regimes~\cite{Shibata95,Anninos96b,Anninos94d}, in preparation for the
study of fully {\it nonlinear, strong field } 3D wave dynamics.
However, until now no such studies have been successfully carried out.

In this paper we present the first successful 
simulations of highly nonlinear gravitational waves in 3D. We study
the process of strong waves collapsing to form BHs under their own
self-gravity.  We determine the critical amplitude for the formation
of BHs and show that one can now carry out these evolutions for long
times.  For waves that are not strong enough to form BHs, we follow
their implosion, bounce and dispersal.  For waves strong enough to
collapse to a BH under their own self gravity, we find the dynamically
formed apparent horizons (AHs), and extract the gravitational
radiation generated in the collapse process.  These waveforms can be
compared in axisymmetry to head-on BH collisions (performed earlier
and reported in~\cite{Anninos94b}). The waveforms are
similar at late times, dominated by the quasi-normal modes of the
resulting BHs as expected.  The difference in the waveforms at early
times for these two very different collapse scenarios shows to what
extent one can extract information about the BH formation process from
the observation of the gravitational radiation emitted by the system.
All the simulations presented here were performed with the newly
developed Cactus code.  For a description of the code and the
numerical methods used,
see~\cite{Bona98b,Alcubierre98a,Seidel98c,Allen98a,Allen99a}.

We take as initial data a pure Brill~\cite{Brill59} type gravitational 
wave, later studied by Eppley~\cite{Eppley77,Eppley79} and 
others~\cite{Holz93}.  The metric takes the form
\begin{equation}
ds^2 = \Psi^4 \left[ e^{2q} \left( d\rho^2 + dz^2 \right) 
+ \rho^2 d\phi^2 \right] =\Psi^4 \hat{ds}^{2},
\label{eqn:brillmetric}
\end{equation}
where $q$ is a free function subject to certain boundary conditions.  
Following~\cite{Allen98a,Camarda97a,Brandt97c}, we choose 
$q$ of the form
\begin{equation}
q = a \; \rho^2 \; e^{-r^2} \left[1 + c \; \frac{\rho^{2}}{(1+\rho^{2})} \;
\cos^{2} \left( n \phi \right) \right],
\end{equation}
where $a,c$ are constants, $r^2 = \rho^2 + z^2$ and $n$ is an integer.
For $c=0$, these data sets reduce to the Holz~\cite{Holz93}
axisymmetric form, recently studied in full 3D Cartesian coordinates
in preparation for the present work~\cite{Alcubierre98b}.  Taking this
form for $q$, we impose the condition of time-symmetry, and solve the
Hamiltonian constraint numerically in Cartesian coordinates.  An
initial data set is thus characterized only by the parameters
$(a,c,n)$.  For the case $(a,0,0)$, we found in~\cite{Alcubierre98b}
that no AH exists in initial data for $a < 11.8$, and we also studied
the appearance of an AH for other values of $c$ and $n$.

Such initial data can be evolved in full 3D using the Cactus code,
which allows the use of different formulations of the Einstein
equations, different coordinate conditions, and different numerical
methods. Our focus here is on new physics, but since stable evolutions
of such strong gravitational waves have not been obtained before, we
comment briefly on the method used for the results in this paper.  In
\cite{Baumgarte99}, Baumgarte and Shapiro note for weak waves that a
system, which is essentially the conformally decomposed ADM system of
Shibata and Nakamura \cite{Shibata95}, shows greatly increased
numerical stability over the standard ADM formulation \cite{York79}.
We will refer to this system as the BSSN formulation. The use of a
particular connection variable in BSSN is reminiscent of the
Bona-Mass\'o formulation \cite{Bona92,Arbona99}.  We found that BSSN
as given in \cite{Baumgarte99} with maximal slicing, a 3-step
iterative Crank-Nicholson (ICN) scheme, and a radiative (Sommerfeld)
boundary condition is very stable and reliable even for the strong
waves considered here. The key new extensions to previous BSSN results
are that the stability can be extended to (i) strong, dynamical fields
and (ii) maximal slicing, where the latter requires some care.
Maximal slicing is defined by vanishing of the mean extrinsic
curvature, $K$=0, and the BSSN formulation allowed us to cleanly
implement this feature numerically, in contrast with the standard ADM
equations. (A related idea to improve stability with maximal slicing
is that of K-drivers, which helps dramatically, but is ultimately not
sufficient for very strong waves in standard ADM formulations
\cite{Balakrishna96a}, but compare \cite{Arbona99}.)  

\begin{figure}
\vspace{-0.5cm}
\epsfxsize=3.4 in
\epsfysize=3.4 in
\epsfbox{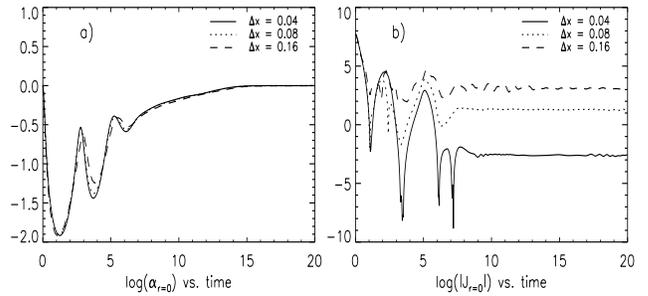} 
\vspace{-3.8cm}
\caption{
  a) Evolution of the log of the lapse $\alpha$ at $r$=0 for the
  axisymmetric data (4,0,0).  The dashed/dotted/solid lines represent
  simulations at low/medium/high resolution.  b) Evolution of the
  Riemann invariant $J$ at $r$=0.  The wave disperses after dynamic
  evolution, leaving flat space behind.}
\label{fig:a4}
\end{figure}

We begin our discussion of the physical results with the parameter set
($a$=4, $c$=0, $n$=0); a rather strong axisymmetric Brill wave (BW).
Even though this data set is axisymmetric, the evolution has been
carried out in full 3D, exploiting the reflection symmetry on the
coordinate planes to evolve only one of the eight octants.  The
evolution of this data set shows that part of the wave propagates
outward while part implodes, re-expanding after passing through the
origin.  However, due to the non-linear self-gravity, not all of it
immediately disperses out to infinity; again part re-collapses and
bounces again.  After a few collapses and bounces the wave completely
disperses out to infinity.  This behavior is shown in
Fig.~\ref{fig:a4}a, where the evolution of the central value of the
lapse is given for simulations with three different grid sizes:
$\Delta x$=$\Delta y$=$\Delta z$=0.16 (low resolution), 0.08 (medium
resolution) and 0.04 (high resolution), using $32^3$, $64^3$ and
$128^3$ grid points respectively.  At late times, the lapse returns to
1 (the log returns to 0).  Fig.~\ref{fig:a4}b shows the evolution of
the log of the central value of the Riemann invariant $J$ for the same
resolutions.  At late times $J$ settles on a constant value that
converges rapidly to zero as we refine the grid. With these results,
and direct verification that the metric functions become stationary at
late times, we conclude that spacetime returns to flat (in non-trivial
spatial coordinates; the metric is decidedly non-flat in appearance!).

\begin{figure}
\vspace{-0.5cm}
\epsfxsize=3.4 in
\epsfysize=3.4 in
\epsfbox{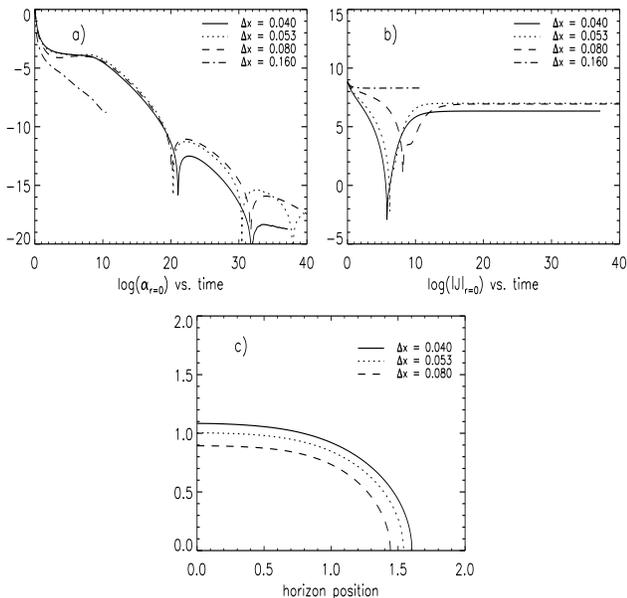}
\caption{
  a) Evolution of the lapse $\alpha$ at $r$=0 for the axisymmetric
  data set (6,0,0).  The dashed, dotted, solid and dashed-dotted lines
  represent simulations at low, medium, intermediate and high resolutions
  respectively.  b) Evolution of the Riemann invariant $J$ at $r$=0.  
  c) Coordinate location of the dynamically formed AH on the $x$-$z$ 
  plane at $t$=10.}
\label{fig:a6}
\end{figure}

Next we increase the amplitude to $a=6$, holding other parameters 
fixed.  Fig.~\ref{fig:a6} shows the evolution of the lapse and the 
Riemann invariant $J$ for this case, showing a clear contrast with 
Fig.~\ref{fig:a4}.  The lapse now collapses immediately, and the 
Riemann invariant after an initial drop grows to a large value at the 
origin until it is halted by the collapse of the lapse.  For this 
amplitude the low resolution is now too crude and the code crashes at 
$t\simeq10$.  We have therefore added an extra simulation with $\Delta 
x$=0.053 (``intermediate'' resolution) using $96^3$ grid points.

To confirm that a BH has indeed formed, we searched for an AH in the
$a$=6 case (using a minimization algorithm~\cite{Alcubierre98b}).  For
high resolution, an AH was first found at $t$=7.7, which grows slowly
in both coordinate radius and area.  Fig.~\ref{fig:a6} shows the
location of the AH on the $x$-$z$ plane at time $t=10$ for the three
resolutions.  The mass of the horizon at this time is about
$M_{AH}=0.87$, but then due to poor resolution of the grid stretching
(a common problem of all BH simulations with singularity avoiding
slicings), it continues to grow, ultimately exceeding the initial ADM
mass of the spacetime, which for this data set is $M_{ADM}=0.99$
(obtained in the way described in~\cite{Alcubierre98b}).
However, the total energy radiated is about 0.12, computed from the
Zerilli functions, completely consistent with $M_{AH}=0.87$ and an
initial mass of $M_{ADM}=0.99$ .  CPU time constraints make it
difficult to run long term, higher resolution simulations (high
resolution used $\sim 120$ hours running on 16 processors of an
SGI/Cray-Origin 2000).  We also confirmed that an event horizon does
not exist in the initial data by integrating null surfaces out from
the origin during the simulation.

From these two studies we conclude that the critical amplitude $a^*$
for BH formation for the axisymmetric BW packet is
\mbox{$a^*=5 \pm 1$}.  We have performed more simulations within this 
range, and have narrowed down the interval to \mbox{$a^*=4.85 \pm 
0.15$}, although near the critical solution higher 
resolution is required to establish convergence.  Our study of these 
near-critical solutions is still under way and will be presented 
elsewhere.

%%%%%%%%%%%%%%%%%%%%%%%%%%%%%%%%%%%%%%%%%%%%%%%%%%%%%%%%%%%%%%%%%%%%

\begin{figure}
\epsfxsize=3.4 in
%\epsfbox{BHcomp_mu1.8.eps} 
%\caption{ We compare the $l$=2,$m$=0 extracted waveform
%for the head-on collision of BHs obtained by
%\protect\cite{Anninos94b} of the $\mu=1.8$ Misner data
%(solid line) to that of the $a=6$ collapsing BW (dotted and
%dashed lines). 
%}
\vspace{-3mm}
\epsfbox{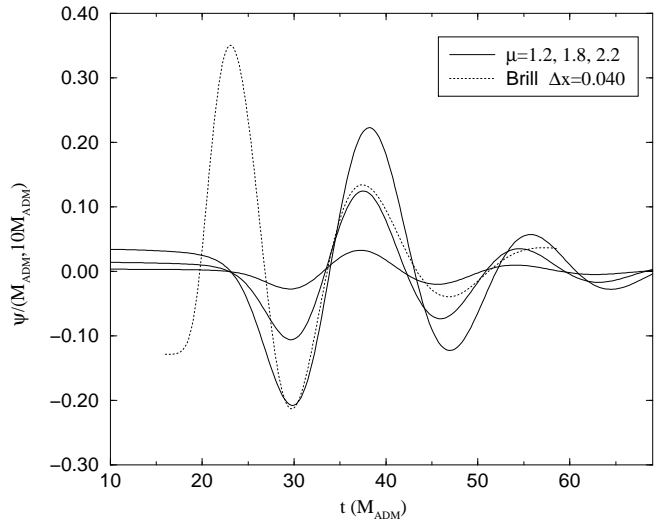}
%\vspace{-3mm}
\caption{ We compare the $l$=2,$m$=0 extracted waveform
for the head-on collision of BHs obtained by
\protect\cite{Anninos94b} of the $\mu=1.2, 1.8, 2.2$ Misner data
(solid lines, increasing amplitude) to that of the $a=6$ collapsing BW 
(dotted and dashed lines). 
}
\label{fig:2Dwave}
\end{figure}

It is particularly exciting that the dynamical evolution can be
followed long enough for the extraction of gravitational waveforms
even for the BH formation case.  One important question is what
physical information of the gravitational collapse process can be
extracted from the observation of the radiation.  How much will the
waveforms from different BH formation processes be different?  For
this purpose we compare the BW collapse waveforms to those of a very
different collapse process, namely the head-on collision of two
BHs. In Fig.~\ref{fig:2Dwave} we show the \{$l$=2,$m$=0\} Zerilli
function $\psi$, obtained from the evolution of Misner data for
$\mu=1.2, 1.8, 2.2$ \cite{Anninos94b}, and from the axisymmetric $a=6$ BW
collapse. (The case $\mu=1.2$ represents a single perturbed black
hole, at $\mu=2.2$ there are two separate black holes that are outside 
the perturbative regime.)
To compare the waveforms, we adjust the time coordinate of the
BW waveforms based on the time delay for different ``detector'' positions,
which for the BW is at $r = 4.6 M_{ADM}$ and for the BHs at $r = 20
M_{ADM}$.  We also scale the Zerilli function amplitude for the BHs by
$M_{ADM}$ and the BW by $10 M_{ADM}$ to put them on the same figure.

We notice the following: 
(1) The BW waveform is dominated by quasi-normal modes (QNMs) at late
times just like in the 2BH case, as expected. 
%The very late part of the BW waveform is still drifting towards convergence. 
A QNM fit shows that at about $10M_{adm}$ from the beginning of the wave-train
the fundamental mode dominates.
(2) However, the BW waveform has more high frequency QNM components in
the early phase.  The waveforms start with a different offset from
zero, which is substantially larger in magnitude in the Brill wave
case, but note that in the BW case the detector is put much closer
in (at $4.6M_{adm}$) and the Zerilli function extraction process
\cite{Abrahams88b,Allen98a} gives a larger ``Coulomb'' component
\cite{Anninos95g}. 
(3) The fundamental QNMs that dominate the late time evolutions for
the two cases have the same phase! We see that both waveforms dip at
$30M_{adm}$, and peak at $38M_{adm}$, to high accuracy.  We note that
the 2BH waveforms for all $\mu \leq 2.2$ have their fundamental QNM
appearing with about the same phase, and we see here the BW collapse
case also has the same phase.  This and other interesting comparisons
between the two collapse scenarios will be discussed further
elsewhere.  (We note that these features noted above are not sensitive
to the $\mu$ value chosen, within the range of $\mu=1.2-2.2$.)

%%%%%%%%%%%%%%%%%%%%%%%%%%%%%%%%%%%%%%%%%%%%%%%%%%%%%%%%%%%%%%%%%%%%%%%%

\begin{figure} 
\epsfxsize=3.4 in
\epsfysize=3.4 in
\vspace{-2mm}
\epsfbox{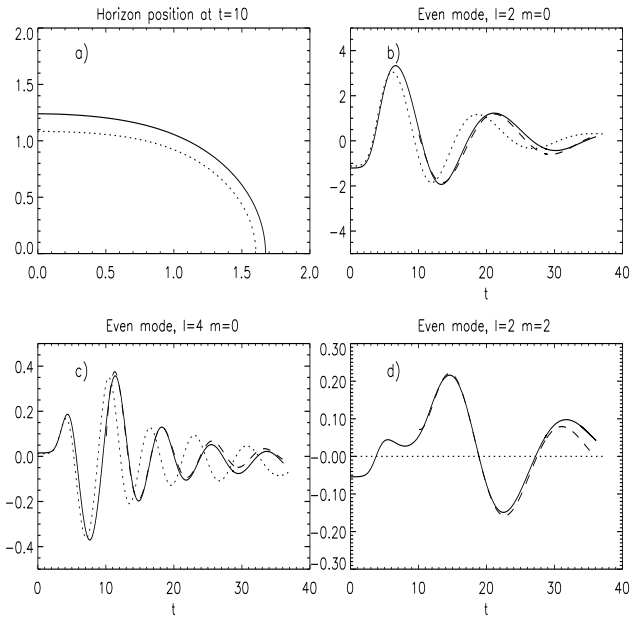}
\caption{a) The solid (dotted) line is the AH for the full 3D data set 
$(6,0.2,1)$ ($(6,0,0)$) at $t$=10 on the $x$-$z$ plane.  b) The
\{$l$=2,$m$=0\} waveform for the 3D $(6,0.2,1)$ case at $r=4$ (solid
line) is compared to axisymmetric $(6,0,0)$ case (dotted line). The
dashed line shows the fit of the 3D case to the corresponding mode
for a BH of mass 1.0.  c) Same comparison for the \{$l$=4,$m$=0\}
waveform. d) Same comparison for the non-axisymmetric \{$l$=2,$m$=2\}
waveform.}
\label{fig:3D}
\end{figure}

Next we go to a pure strong wave case with full 3D features (the first
ever simulated), where the initial waveform is even more dominated by
details of the BH formation process.  Fig.~\ref{fig:3D} shows
the development of the data set ($a$=6, $c$=0.2, $n$=1), which 
has reflection symmetry across coordinate planes; it again suffices to
evolve only an octant.  The initial ADM mass of this data set turns
out to be $M_{ADM}=1.12$.  Fig.~\ref{fig:3D}a shows a comparison of
the AHs of this 3D and the previous axisymmetric cases, using the same
high resolution, at $t$=10 on the $x$-$z$ plane.  The mass of the 3D
AH case is larger, weighing in at $M_{AH}$=0.99 (compared to
$M_{AH}(2D)=0.87$).

In Fig.~\ref{fig:3D}b we show the \{$l$=2,$m$=0\} waveform of this 3D
case, compared to the previous axisymmetric case.  The $c=0.2$
waveform has a longer wave length at late times, consistent with the
fact that a larger mass BH is formed in the 3D case.
Figs.~\ref{fig:3D}c and~\ref{fig:3D}d show the same comparison for the
\{$l$=4,$m$=0\} and \{$l$=2,$m$=2\} modes respectively.  Notice that
while the first two modes are of similar amplitude for both runs, the
3D \{$l$=2,$m$=2\} mode is completely different; as a non-axisymmetric
contribution, it is absent in the axisymmetric run (in fact, it
doesn't quite vanish due to numerical error, but it remains of order
$10^{-6}$).  We also show a fit to the corresponding QNM's of a BH of
mass 1.0.  The fit was performed in the time interval $(10,36)$, and
is noticeably worse if the fit is attempted to earlier times, again
showing that the lowest QNM's dominate at around $10$.  The
early parts of the waveforms $t<10$ reflect the details of the initial
data and BH formation process.  This is especially clear in the
\{$l$=2,$m$=2\} mode, which seems to provide the most information
about the initial data and the 3D BH formation process.  At present no
3D BH formation simulation from other scenarios (e.g., true spiraling
BH coalescence) are available for comparison, as in the axisymmetric
case, but such simulations may actually be available
soon~\cite{Bruegmann97}. It will be interesting to compare such
studies with 3D wave collapses, such as that presented here.

In conclusion, we demonstrated numerical evolutions of 3D, strongly
non-linear gravitational waves, and studied gravitational collapse of
axisymmetric {\it and} non-axisymmetric gravitational waves.  We
compared the wave collapse to the head-on collision of two black
holes. The research opens the door to many investigations.

\noindent {\bf Acknowledgments.} This work was supported by AEI, NCSA, 
NSF PHY 9600507, NSF MCA93S025 and NASA NCCS5-153.  We thank many 
colleagues at the AEI, Washington University, Univeristat de les Illes 
Balears, and NCSA for the co-development of the Cactus code, and 
especially D. Holz for important discussions.  
Calculations were performed at AEI, NCSA, SDSC, RZG-Garching and ZIB 
in Berlin.

\bibliographystyle{prsty}
%\bibliography{bibtex/references}

\begin{thebibliography}{10}

\bibitem{Wheeler55}
J.~A. Wheeler, Phys. Rev. {\bf 97},  511  (1955).

\bibitem{Brill64}
D. Brill and J. Hartle, Phys. Rev. {\bf 135},  B271  (1964).

\bibitem{Yurtsever88b}
U. Yurtsever, Phys. Rev. D {\bf 38},  1731  (1988).

\bibitem{Abrahams92b}
A. Abrahams and C. Evans, Phys. Rev. D {\bf 46},  R4117  (1992).

\bibitem{Abrahams93a}
A. Abrahams and C. Evans, Phys. Rev. Lett. {\bf 70},  2980  (1993).

\bibitem{Gundlach97b}
C. Gundlach,   (1998), gr-qc/9710066.

\bibitem{Shibata95}
M. Shibata and T. Nakamura, Phys. Rev. D {\bf 52},  5428  (1995).

\bibitem{Anninos96b}
P. Anninos {\it et~al.}, Phys. Rev. D {\bf 54},  6544  (1996).

\bibitem{Anninos94d}
P. Anninos {\it et~al.}, Phys. Rev. D {\bf 56},  842  (1997).

\bibitem{Anninos94b}
P. Anninos {\it et~al.}, Phys. Rev. D {\bf 52},  2044  (1995).

\bibitem{Bona98b}
C. Bona, J. Mass{\'o}, E. Seidel, and P. Walker,   (1998), gr-qc/9804065.
  Submitted to Physical Review D.

\bibitem{Alcubierre98a}
M. Alcubierre {\it et~al.},   (1998), in preparation.

\bibitem{Seidel98c}
E. Seidel and W.-M. Suen, J. Comp. Appl. Math.  (1999), in press.

\bibitem{Allen98a}
G. Allen, K. Camarda, and E. Seidel,   (1998), gr-qc/9806036. Submitted to
  Phys. Rev. D.

\bibitem{Allen99a}
G. Allen, T. Goodale, and E. Seidel,  in {\em 7th Symposium on the Frontiers of
  Massively Parallel Computation-Frontiers 99} (IEEE, New York, 1999).

\bibitem{Brill59}
D.~S. Brill, Ann. Phys. {\bf 7},  466  (1959).

\bibitem{Eppley77}
K. Eppley, Phys. Rev. D {\bf 16},  1609  (1977).

\bibitem{Eppley79}
K. Eppley,  in {\em Sources of Gravitational Radiation}, edited by L. Smarr
  (Cambridge University Press, Cambridge, England, 1979), p.\ 275.

\bibitem{Holz93}
D. Holz, W. Miller, M. Wakano, and J. Wheeler,  in {\em Directions in General
  Relativity: Proceedings of the 1993 International Symposium, Maryland; Papers
  in honor of Dieter Brill}, edited by B. Hu and T. Jacobson (Cambridge
  University Press, Cambridge, England, 1993).

\bibitem{Camarda97a}
K. Camarda, Ph.D. thesis, University of Illinois at Urbana-Champaign, Urbana,
  Illinois, 1998.

\bibitem{Brandt97c}
S. Brandt, K. Camarda, and E. Seidel,  in {\em Proc. 8th M. Grossmann Meeting},
  edited by T. Piran (World Scientific, Singapore, 1998), in press.

\bibitem{Alcubierre98b}
M. Alcubierre {\it et~al.},   (1998), gr-qc/9809004.

\bibitem{Baumgarte99}
T. Baumgarte and S. Shapiro, Physical Review D {\bf 59},  024007  (1999).

\bibitem{York79}
J. York,  in {\em Sources of Gravitational Radiation}, edited by L. Smarr
  (Cambridge University Press, Cambridge, England, 1979).

\bibitem{Bona92}
C. Bona and J. Mass\'{o}, Phys. Rev. Lett. {\bf 68},  1097  (1992).

\bibitem{Arbona99}
A. Arbona, C. Bona, J. Mass{\'o}, and J. Stela, gr-qc/9902053  (1999).

\bibitem{Balakrishna96a}
J. Balakrishna {\it et~al.}, Class. Quant. Grav. {\bf 13},  L135  (1996).

\bibitem{Abrahams88b}
A. Abrahams and C. Evans, Phys. Rev. D {\bf 37},  318  (1988).

\bibitem{Anninos95g}
P. Anninos {\it et~al.}, Phys. Rev. D {\bf 52},  4462  (1995).

\bibitem{Bruegmann97}
B. Br{\"u}gmann, Int. J. Mod. Phys. D {\bf 8},  85  (1999).

\end{thebibliography}

\end{document}